\documentclass[journal=jacsat,manuscript=article]{achemso}

\usepackage{chemformula} 
\usepackage[T1]{fontenc} 

\author{Weijian Li}
\affiliation[APP Rice]
{Applied Physics Graduate Program, Smalley-Curl Institute, Rice University, Houston, TX 77005, USA}
\author{Gururaj V. Naik}
\affiliation[ECE Rice]
{Electrical \& Computer Engineering, Rice University, TX 77005, USA}
\email{guru@rice.edu}
\phone{+1 (713)-348-2528}

\title{Large optical tunability from charge density waves in 1T-TaS$_2$ under incoherent illumination}

\abbreviations{CDW,TMDs}

\begin{document}


\begin{abstract}

  Strongly correlated materials possess a complex energy landscape and host many interesting physical phenomena, including charge density waves (CDWs). CDWs have been observed and extensively studied in many materials since their first discovery in 1972. Yet, they present ample opportunities for discovery. Here, we report a large tunability in the optical response of a quasi-2D CDW material, 1T-TaS$_2$, upon incoherent light illumination at room temperature. We show that the observed tunability is a consequence of light-induced rearrangement of CDW stacking across the layers of 1T-TaS$_2$. Our model, based on this hypothesis, agrees reasonably well with experiments suggesting that the interdomain CDW interaction is a vital knob to control the phase of strongly correlated materials.

\end{abstract}

\section{Introduction}

  Light-matter interaction in strongly correlated materials is interesting because light can significantly alter the free energy landscape of strong correlations resulting in many new physical phenomena \cite{matsunaga2014light,sie2019ultrafast,tao2016nature}. Optical excitation can tip the balance between various competing forms of order leading to photo-induced metal-to-insulator transitions \cite{hu2016broadband,lourembam2016evidence,perfetti2006time}, charge density waves \cite{vaskivskyi2015controlling,frigge2017optically,perfetti2008femtosecond,stahl2020Collapse}, ferromagnetic/antiferromagnetic transitions \cite{eggebrecht2017light,stupakiewicz2017ultrafast}, superconductivity \cite{takasan2017laser}, and others \cite{ohkoshi2010synthesis}. Though light-induced changes in the lattice, electrical, and magnetic properties have been extensively studied before, optical properties remain much to be explored. In this work, we study the tunable optical properties of a strongly correlated material, 1T-TaS$_2$, which supports charge density waves (CDWs) at room temperature.

  Many chalcogenides and organic compounds support CDWs at low temperatures \cite{gruner1988dynamics}. However, 1T-TaS$_2$, 1T-TaSe$_2$, and a few other lanthanide tellurides support CDWs at room temperature and are interesting for device applications \cite{hovden2016atomic,wu1990direct,bovet2004pseudogapped,slough1990atomic,burk1991charge}. Among them, 1T-TaS$_2$ is the only material that exhibits nearly commensurate CDW (NCCDW) phase at room temperature resulting in a large tunability in its electrical conductivity. 

  The tunability of electrical properties of 1T-TaS$_2$ is extensively studied in the past. 1T-TaS$_2$ was shown to exhibit nonlinear electrical conductance at room temperature \cite{di1977low}, and hysteresis behavior of its electric resistance \cite{uchida1978nonlinear,zhu2018light,stojchevska2018stability}. Also, the CDW phase transition was demonstrated to be sensitive to pressure \cite{ravy2012high,sipos2008mott}, strain \cite{zhao2017tuning,svetin2014transitions}, thickness \cite{yoshida2015memristive,fu2016controlled}, gate voltage \cite{hollander2015electrically,yu2015gate,geremew2019bias}, and chemical doping \cite{li2012fe,rossnagel2010suppression}. However, the optical properties of 1T-TaS$_2$ and its tunability remain unexplored. In this work, we study light-tunable optical properties of 1T-TaS$_2$.

  The charge order in 1T-TaS$_2$ manifests as a lattice reorganization where 12 Ta atoms surrounding a central Ta atom move slightly inwards forming a David-star structure (see Fig. \ref{fig1}a). Groups of such David-stars called CDW domains exist in each layer of 1T-TaS$_2$ at room temperature. The relative position or stacking arrangement of such CDW domains across layers has been recently reported to impact the electronic bandstructure of 1T-TaS$_2$ significantly \cite{ritschel2018stacking,ritschel2015orbital,ma2016metallic,le2017stacking}. Depending on the stacking, the material can be insulating or metallic along its c-axis. Such drastic dependence of the electronic properties on the CDW stacking configuration is unusual and forms the central theme of this work. Here, we observe a substantial change in the dielectric function of 1T-TaS$_2$ only in the c-axis upon incoherent white light illumination. We attribute the observed changes in the dielectric function to the stacking rearrangement of the CDW domains in 1T-TaS$_2$.

\section{Results and discussion}

\begin{figure}[htbp]
\centering\includegraphics[width = 15.6cm]{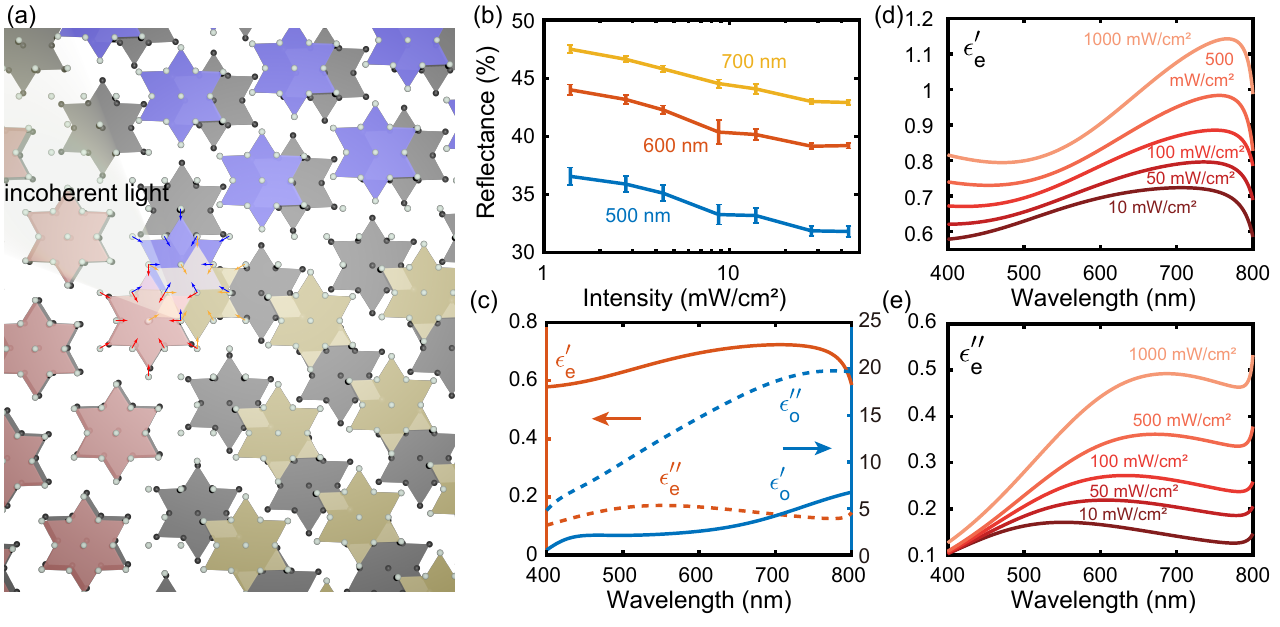}
\caption{Optical properties of 1T-TaS$_2$ illuminated by white light at room temperature: (a) A schematic showing the CDW domains in two adjacent layers of 1T-TaS$_2$. The David stars in the bottom (gray), and top (colored) layers can stack in different configurations, including center-to-center (red), center-to-shoulder (blue) and center-to-corner (yellow) stacking. The stacking configuration can switch with incoherent light, as indicated by the arrows at the center of the domain. (b) The reflectance of a 1T-TaS$_2$ thin film measured at three different wavelengths as a function of incident white light intensity. (c) Real (solid) and imaginary (dashed) components of ordinary ($\varepsilon_o$) and extraordinary or c-axis ($\varepsilon_e$) permittivity of 1T-TaS$_2$ retrieved from low intensity optical measurements. Only $\varepsilon_e$ shows white light tunability. (d) Real and (e) imaginary components of $\varepsilon_e$ at different intensities of white light.}
\label{fig1}
\end{figure}

  We obtained crystalline 1T-TaS$_2$ thin film samples by mechanical exfoliation, as reported in our previous work \cite{li2019plane}. The reflectance of a 180 nm thick film measured at three different wavelengths using an objective of 0.85 numerical aperture is shown in Fig. \ref{fig1}b as a function of illumination intensity. The 1T-TaS$_2$ film exhibits an intensity-dependent reflectance even at low illumination intensities of a few mW/cm$^2$. For comparison, the peak intensity of 1.5AM sun is approximately 100 mW/cm$^2$.
  
  Using a Fourier plane imaging spectrometer, we measure the reflection and transmission spectra of the 1T-TaS$_2$ film at incident angles ranging from 0 to 53$^\circ$. From this data, we extract the anisotropic dielectric function of 1T-TaS$_2$ in the visible (see SI). The in-plane and out-of-plane (along c-axis) dielectric functions denoted by $\varepsilon_o$ and $\varepsilon_e$ respectively are shown in Fig. \ref{fig1}c for low illumination. 1T-TaS$_2$ is strongly anisotropic and highly absorbing dielectric in the visible.

  As we increase the incoherent illumination intensity up to 1000 mW/cm$^2$, the in-plane dielectric function remains the same (see SI), but the out-of-plane dielectric function changes. The real and imaginary parts of the intensity-dependent $\varepsilon_e$ are as shown in Fig. \ref{fig1}d and e. The real component of out-of-plane epsilon ($\varepsilon_e^{'}$) increases by about 0.4 and the imaginary component ($\varepsilon_e^{''}$) increases by about 0.5 as the incident incoherent light intensity increases from 10 mW/cm$^2$ to 1000 mW/cm$^2$. 

  The observed light-induced change in c-axis dielectric function could be a result of many possible microscopic phenomena. The low intensity of the white light used in these experiments rule out photothermal effects (see SI). Light-induced free carrier generation could account for the observed change in optical constants only if the lifetime of these carriers is about a millisecond. Such a large lifetime for free carriers is not physical. Additionally, the dielectric constants changing only along c-axis suggests a mechanism that involves interlayer interaction. To further probe the underlying mechanism, we carry out time-resolved measurements.
  
\begin{figure}[htbp]
\centering\includegraphics[width=9cm]{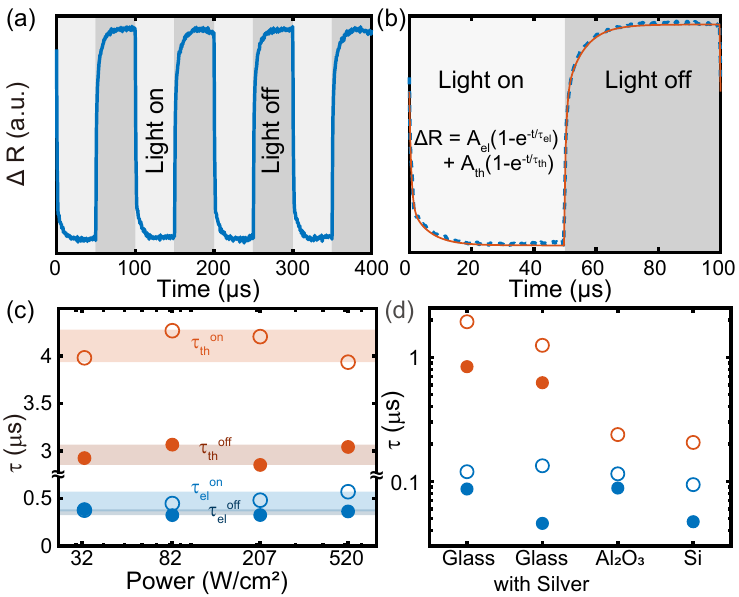}
\caption{Time-resolved measurements of reflectance of 1T-TaS$_2$ under a square-wave modulated pump laser at 640 nm wavelength: (a) The change in reflectance ($\Delta R$) at 785 nm wavelength recorded for multiple cycles. (b) A two-exponential function fit to the measured $\Delta R$ at each edge of a cycle is plotted. (c) The extracted relaxation time-constants at light-on (hollow circles) and light-off (filled circles) edges as a function of pump intensity at 10 kHz modulation. (d) The time-constants extracted for 1T-TaS$_2$ on four different substrates with different thermal conductivity at 250 kHz modulation.}
\label{fig2}
\end{figure}

  At first, the change in optical properties of 1T-TaS$_2$ with illumination was suspected to arise from free carriers. Hence, a ps-pulsed laser was used to measure time-resolved reflectance (see SI). We observed a $\mu$s response time and hence ruled out a free carrier mechanism for the change in optical constants. Then, we perform time-resolved measurements with a few kHz to a few 100's of kHz modulated laser incident on 1T-TaS$_2$ thin films. The measured change in reflectance of the 1T-TaS$_2$ film at 785 nm wavelength is plotted against time in Fig. \ref{fig2}a. A square wave modulates the pump laser at 640 nm wavelength and the pump light-on and light-off regions are shown in Fig. \ref{fig2}. The modulated reflectance signal from the 1T-TaS$_2$ film was repeatable for at least 600 s in ambient air.

  The observed change in reflectance ($\Delta R$) could be fitted well with a sum of two exponential functions, as indicated in Fig. \ref{fig2}b. The two time-constants each for light-on and light-off transients are plotted in Fig. \ref{fig2}c as a function of the peak pump intensity. The time-constants are nearly independent on intensity and suggest that there must be at least two mechanisms behind the observed $\Delta R$. Since the smallest laser intensity in Fig. \ref{fig2} is about an order of magnitude more than that of white light in Fig. \ref{fig1}, photothermal effects could be expected as one of the mechanisms contributing to the $\Delta R$ in Fig. \ref{fig2}. Many previous studies on 1T-TaS$_2$ have used intense lasers and observed photothermal effects, including phase transition to incommensurate CDW (ICCDW) state \cite{laulhe2017ultrafast,zheng2017room}.

  Photothermal effects depend on the thermal conductivity of the substrate. Hence, measuring time-constants for 1T-TaS$_2$ on different substrates could elicit the photothermal effect. Fig. \ref{fig2}d shows the time-constants measured on four different substrates: glass, glass with an 80 nm silver layer, sapphire, and silicon. One of the two time-constants shrinks with increasing thermal conductivity of the substrate for both light-on and light-off cases. We attribute this time-constant to the photothermal phenomenon. The difference between the photothermal time-constants for light-on and light-off cases could arise from the temperature-dependent thermal conductivity of 1T-TaS$_2$. Note that $\tau_{th}^{off}$ for sapphire and silicon substrates are small and comparable to the other time-constant. Hence, its extraction was not uniquely possible.

  While the photothermal time-constant shows a strong substrate dependence, the other does not. This other time-constant could be related to the CDW phenomenon because CDWs in 1T-TaS$_2$ typically exhibit a response time of 100's of ns. This underlying CDW phenomenon should be common to both measurements, under white light (Fig. \ref{fig1}) and laser (Fig. \ref{fig2}), and account for the observed light-induced change in c-axis dielectric function. 

  Since dielectric function changes only along the c-axis, light could be rearranging CDW domains in the layers of 1T-TaS$_2$ resulting in a change in CDW stacking across layers. Recently, CDW stacking has been shown to strongly influence the band structure of 1T-TaS$_2$, especially in the $\Gamma-$A direction \cite{ritschel2018stacking}. Thus, we hypothesize that the change in the c-axis dielectric function of 1T-TaS$_2$ originates from the rearrangement of CDW stacking across layers. 
  
\begin{figure}[htbp]
\centering\includegraphics[width=9cm]{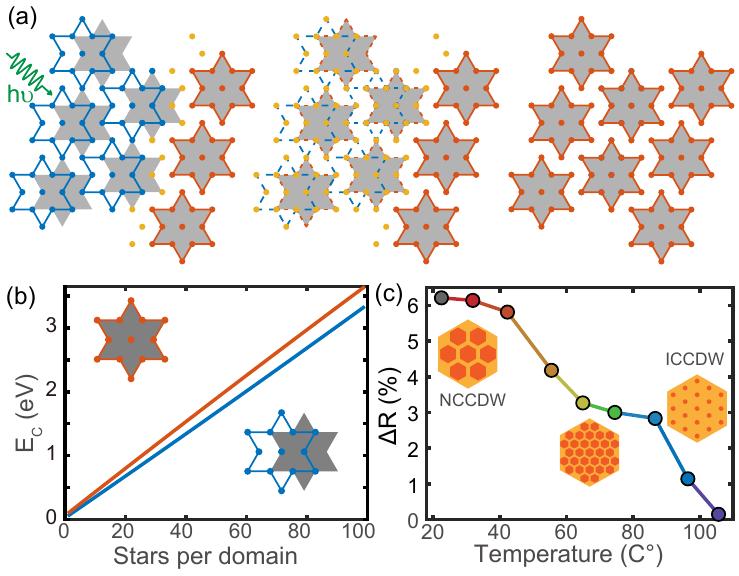}
\caption{The mechanism of light-induced CDW stacking rearrangement: (a) Schematic showing two possible stacking configurations in two adjacent layers of 1T-TaS$_2$. The gray David stars make the bottom layer. Blue and red stars correspond to center-to-shoulder and center-to-center stacking, respectively. Light absorption in a blue domain breaks the formation and can eventually lead to the formation of a red domain. (b) Calculated interlayer Coulomb energy per David-star vs. domain size for the two stacking types shown in (a). (c) Measured reflectance change of 1T-TaS$_2$ film at 700 nm wavelength as a function of temperature. The inset corresponds to the typical CDW domains in NCCDW and ICCDW phases.}
\label{fig3}
\end{figure}

  The proposed mechanism of light-induced rearrangement of CDW stacking is depicted in the schematic of Fig. \ref{fig3}a. At first, the absorption of a visible light photon breaks a CDW domain. Subsequently, new domains with different stacking configurations appear. The schematic identifies the most likely stacking configurations with and without light. In the dark, the Coulomb repulsion between the CDW domains energetically favors the center-to-shoulder stacking configuration. Upon illumination, some David stars stack on top of each other in the center-to-center configuration. The calculated interlayer Coulomb repulsion energies for the two stacking configurations are shown in Fig. \ref{fig3}b as a function of domain size (See SI for further details of calculations). At room temperature, a typical domain contains a few tens of stars \cite{tsen2015structure}, and hence the energy difference between the two stacking types is about 150 meV. As the temperature increases, the domain size shrinks and reduces the energy difference between the two stacking configurations. Thus, increasing temperature makes both stacking configurations more equally probable. Therefore, the tunability of the c-axis dielectric function should diminish with increasing temperature and eventually vanish in the ICCDW phase. Fig. \ref{fig3}c plots the temperature-dependent $\Delta R$ measured at 700 nm wavelength with a low intensity incoherent optical pump. The tunability drops with increasing temperature and vanishes in the ICCDW phase as expected. 

\begin{figure}[htbp]
\centering\includegraphics[width=9cm]{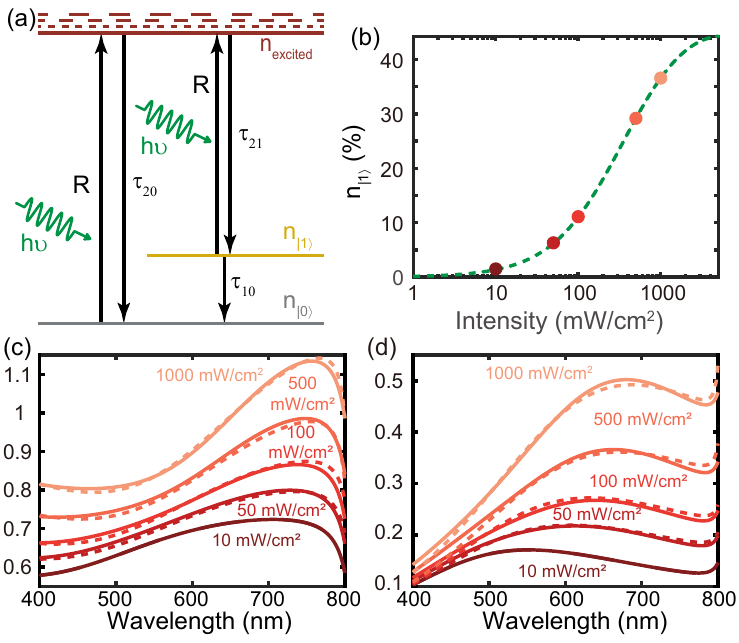}
\caption{Tunability in c-axis dielectric function of 1T-TaS$_2$ from CDW stacking rearrangement: (a) A three-level system to model the light-induced CDW stacking rearrangement. $|0\rangle$ and $|1\rangle$ correspond to the two stacking configurations. The excited state corresponds to the free-electron energy band. \textit{R} is the light-induced transition rate for either of the stacking to the excited state. All $\tau$ correspond to the respective relaxation time-constants. (b) Calculated population fraction in $|1\rangle$ stacking type as a function of excitation intensity (\textit{R}). Red dots correspond to intensity values used in experiments. Real (c) and imaginary (d) components of the c-axis dielectric function for different excitation intensity. Theoretical (solid) and experimental (dashed) curves are plotted in both panels.}
\label{fig4}
\end{figure}

  The proposed mechanism of CDW stacking rearrangement involves at least three steps: setting electrons and phonons of a domain free by optical absorption, formation of a new domain in center-to-center stacking configuration (stacking $|1\rangle$) and relaxation to ground state or center-to-shoulder stacking configuration (stacking $|0\rangle$). The three processes together could be modeled as a three-level system shown in Fig. \ref{fig4}a. The relaxation times for the formation of a new domain in stacking configurations $|0\rangle$ and $|1\rangle$ are $\tau_{20}$ and $\tau_{21}$ respectively. The relaxation time for a transition from stacking $|1\rangle$ to $|0\rangle$ is $\tau_{10}$. $\tau_{10}$ should correspond to the light-off relaxation time from our time-resolved measurements ($\tau_{el}$). $\tau_{20}$ and $\tau_{21}$ do not involve any stacking change and hence should be much smaller than $\tau_{10}$. Previous works report $\tau_{20}$ and $\tau_{21}$ in the order of 500 ps around 200 K, and show an increasing trend with higher temperature \cite{laulhe2017ultrafast,vaskivskyi2015controlling}. Therefore, we expect $\tau_{20}$ and $\tau_{21}$ to be a few nanoseconds at room temperature.

  Using the three-level model, the steady-state fractions of CDW domains in the two stacking configurations could be computed as a function of illumination intensity. Fig. \ref{fig4}b plots the fraction of CDW domains in stacking $|1\rangle$ ($n_1$) as a function of incident light intensity. Note that the equilibrium value of $n_1$ is approximated to zero because the energy gap between the two staking levels is more than $5k_BT$ at room temperature (previous discussion about Fig. \ref{fig3}b). From the rate equations, at steady-state, $n_1=\frac{\tau_{10}}{\tau_{21}(1+R\tau_{10})}\frac{R\tau_{20}\tau_{21}}{R\tau_{20}\tau_{21}+\tau{21}+\tau{20}}$. In the limit, $\tau_{10}>>\tau_{21}$ and $\tau_{10}>>\tau_{20}$, $\frac{n_1}{N}\approx\frac{R\tau_{10}}{2(1+R\tau_{10})}$ for $R<<1/\tau_{21}$. We set $\tau_{21}=\tau_{20}=$3 ns and $\tau_{10}=$0.38 $\mu$s. For the intensities used in this work, $R<<1/\tau_{21}$ and thus the maximum value of $n_1$ is close to 50\%. Also, $n_1$ as a function of $R$ is determined by only one parameter, $\tau_{10}$. Since, our time-resolved experiments measured $\tau_{10}$, we expect the prediction of Fig. \ref{fig4}b to be quantitative.

  The electronic bandstructure of 1T-TaS$_2$ along $\Gamma-A$ has been predicted to depend on the CDW stacking order \cite{ritschel2018stacking}. Hence, we associate each stacking configuration to a unique dielectric function along the c-axis. We calculate the c-axis dielectric function for any intensity of incident light by using the population fraction information from Fig. \ref{fig4}b in Maxwell-Garnet effective medium equation \cite{levy1997maxwell}. The lowest intensity of incoherent light used in Fig. \ref{fig1} corresponds to $n_1\approx$1\% and hence may be considered as the c-axis dielectric function of ground state stacking. Since this ground state $\varepsilon_e$ is a two-Lorentzian curve, we expect the $\varepsilon_e$ corresponding to stacking $|1\rangle$ is also a two-Lorentzian curve. The parameters of the two Lorentzian oscillators corresponding to stacking $|1\rangle$ are obtained by fitting the effective permittivity from the Maxwell-Garnett equation to the measured dielectric function for the highest intensity, 1000 mW/cm$^2$. The value of $n_1$ for this intensity is taken from Fig. \ref{fig4}b. The $\varepsilon_e$ corresponding to the intermediate three intensities are calculated from the Maxwell-Garnett equation incorporating appropriate values of $n_1$ shown in Fig. \ref{fig4}b. The experimental and calculated c-axis dielectric functions for various light intensities are plotted in Fig. \ref{fig4}c, and Fig. \ref{fig4}c. The effective medium calculations agree well with experiments providing quantitative evidence for our hypothesis of light-induced CDW stacking rearrangement in 1T-TaS$_2$.

  In conclusion, we discovered a new optical phenomenon in 1T-TaS$_2$. Our work showed that low-intensity white light illumination could change the c-axis permittivity of 1T-TaS$_2$ by unity-order. Time-resolved measurements showed that the speed of change is a few MHz. We hypothesized that the stacking of CDW domains across layers could rearrange with illumination and thereby lead to a change in c-axis optical constants. Our modeling results agreed well with experimental measurements upholding our hypothesis. This discovery proves that the electronic bandstructure depends on the CDW stacking order, and the stacking order may be controlled by light. Our work shows that stacking order is a new dimension in the phase diagram of strongly correlated materials and enriches opportunities for discovery. This work could lead to further development of low-power and fast-tunable optical materials and potentially revolutionize future imaging, display, and sensing applications.

\begin{acknowledgement}

The authors thank Bharadwaj group at Rice University for their help in optical characterization, and also thank all Naik group members for useful discussion and help with preparing this manuscript.

\end{acknowledgement}

\begin{suppinfo}

  The supporting information is available free of charge via the Internet at http://pubs.acs.org.

\begin{itemize}
  \item SI: Methods; Ordinary and extraordinary dielectric functions; Dielectric function extraction; Time-resolved measurements using an ultrafast laser; Photothermal effect; Coulomb energy.
\end{itemize}
\end{suppinfo}

\bibliography{achemso-demo}

\end{document}